\documentclass[aps,reprint,floatfix,twocolumn,prl,superscriptaddress]{revtex4-1}

\usepackage{xcolor} 
\usepackage{graphicx}
\usepackage[hidelinks]{hyperref}
\usepackage{amssymb}
\usepackage[utf8]{inputenc}
\usepackage{amsmath}
\usepackage[normalem]{ulem}
\newcommand{\abs}[1]{\left|#1 \right|}
\newcommand{\mat}[1]{\mathrm{#1}}
\newcommand{\op}[1]{\mathrm{#1}}
\renewcommand{\vec}[1]{\boldsymbol{#1}}

\usepackage{mathtools}
\newcommand{\change}[1]{\textcolor{black}{#1}}

\begin{document}

\title{Quantum buckling in metal-organic framework materials}
\author{R. Matthias Geilhufe}
\affiliation{Nordita,  KTH Royal Institute of Technology and Stockholm University, Roslagstullsbacken 23,  10691 Stockholm,  Sweden}
\email{matthias.geilhufe@su.se}
\date{\today}

\begin{abstract}
Metal organic frameworks are porous materials composed of metal ions or clusters coordinated by organic molecules. As a response to applied uniaxial pressure, molecules of straight shape in the framework start to buckle. Under sufficiently low temperatures, this buckling is of quantum nature, described by a superposition of degenerate buckling states. Buckling states of adjacent molecules couple in a transverse Ising type behavior. On the example of the metal organic framework topology MOF-5 we derive the phase diagram under applied strain, showing a normal, a parabuckling, and a ferrobuckling phase. At zero temperature, quantum phase transitions between the three phases can be induced by strain. This novel type of order opens a new path towards strain induced quantum phases.
\end{abstract}
\maketitle
Under sufficient axial load, a column responds by a sudden deformation - buckling. The deformation corresponds to a classical solution minimizing the action. In the nanoscale, the electrostatic control of buckling was recently realized, giving rise to buckling bits for nanomechanical computation \cite{nanobuckling}. Decreasing the column  size even further, quantum effects become dominant, allowing for the tunneling between degenerate buckling states. Recently, this line of thought has initiated research in realizing mechanical qubits \cite{savel2006quantum,savel2007quantum}. \change{Prominent designs are proposed, e.g., based on carbon nanotubes \cite{qubit1}.} Realizing entanglement between various adjacent mechanical qubits has remained an open question. 

Metal organic framework materials are compounds built of metal ions or clusters coordinated by organic ligands. After the first MOFs were realized in the late 1990s \cite{li1999design}, more than 90,000 stable structures have been synthesized and characterized to date \cite{moosavi2020understanding}. They are intensively discussed in the context of gas sorption and storage, catalysis, electronic devices, etc \cite{mofsrev1,mofsrev2,mofsrev3}. Mechanical properties and flexibility of MOFs are summarized in Ref. \cite{schneemann2014flexible}.

In the present paper it is shown how uniaxial pressure in MOFs can be used to induce quantum buckling of ligand molecules. Interestingly, the buckling of individual molecules is not independent, but interacts similarly to a transverse field Ising model. We motivate the model on the example of molecular buckling in the system MOF-5. Depending on applied uniaxial pressure, the system undergoes two quantum phase transitions: first, from a normal phase to a parabuckling phase; second: from a parabuckling phase to a ferrobuckling phase. In the parabuckling phase, elementary excitations exhibit a gap (mass) of $\Delta = 4\sqrt{t(t-4J)}$. At the parabuckling-ferrobuckling phase transition ($t=4J$) the gap closes. The parabuckling-ferrobuckling quantum critical point might give rise to novel types of fluctuation-induced order, such as strain-controlled superconductivity.

\begin{figure}
    \centering
    \includegraphics[width=0.45\textwidth]{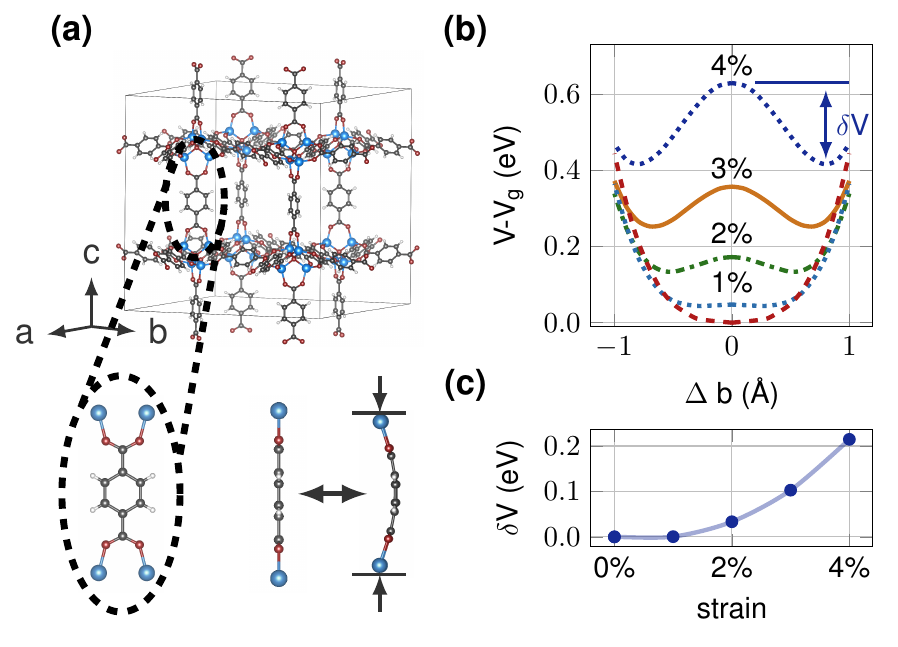}
    \caption{Buckling of BDC molecules in MOF-5. (a) shows the crystal structure of MOF-5. Zn$_4$O clusters are coordinated by BDC molecules, forming a periodic 3D cubic network. BDC Molecules can be buckled by uniaxial pressure. (b) Total energy $V$ versus buckling of BDC molecules for various strain values. $V_g$ denotes the equilibrium total energy. Under strain, the total energy follows the form of a double-well potential. (c) Barrier height $\delta V$ of the double-well potential against uniaxial strain along the Cartesian $z$-axis. }
    \label{StrucEn}
\end{figure}
MOF-5 describes a cubic framework topology \cite{li1999design} with the sum formula Zn$_4$O(BDC)$_3$. BDC is an abbreviation for 1,4-benzenedicarboxylate (Fig. \ref{StrucEn}(a)). The structure is open, with a spacing of 12.94 \AA~between the centers of adjacent Zn$_4$O clusters. MOF-5 is extremely soft with a bulk modulus of 15.37~GPa \cite{yang2010theoretical}. Applying strain along one of the Cartesian axes, the bond lengths within the Zn$_4$O clusters and BDC molecules are squeezed up to the point where the structure responds by a buckling of the molecules. To approximate the effect, we monitor the change in total energy upon buckling, for individual BDC molecules bound to Zn atoms and applied strain along the Cartesian $z$-direction. Calculations based on density functional theory were performed using VASP \cite{vasp}. The exchange-correlation functional was approximated by the MetaGGA functional SCAN \cite{SunSCAN} in combination with the rVV10 Van-der-Waals corrections \cite{SabatinirVV10,SCANVdW}. The energy cutoff was set to 700~eV. Zn-$d$-electrons were treated with a Hubbard-U correction of 2~eV \cite{Dudarev1998}.

The buckling potential for various strain strength is shown in Fig. \ref{StrucEn}(b). By applying strain, the total energy increases and the energy profile forms a double-well potential. The potential has two degenerate minima, the left- and right-buckled state. To lowest order, the potential can be described by a fourth order polynomial,
\begin{equation}
    V(b) = V_0 + a(b^2-b_0^2)^2.
\end{equation}
$a$ is an overall scaling factor and $b_0$ the potential minimum or classical solution. $V_0$ is the characteristic energy. It coincides with the potential minimum for the zero-strain case ($b_0=0$). Note, that all three parameters are strain dependent (see Table \ref{fitting_parm}). 

\begin{table}[]
    \centering
    \begin{tabular}{cccc}
    \hline\hline
     & $V_0$~[eV] & $a$~[eV\AA$^{-4}$] & $b_0$~[\AA] \\
    \hline
      1\% & -171.70 & 0.35 & 0.19 \\
      2\% & -171.61 & 0.40 & 0.54 \\
      3\% & -171.49 & 0.45 & 0.69 \\
      4\% & -171.33 & 0.51 & 0.81 \\
      \hline\hline
    \end{tabular}
    \caption{Fitting parameters for the double-well potential. The table shows the strain dependence of the characteristic energy $V_0$, the scaling factor $a$, and the classical solution or potential minimum $b_0$.}
    \label{fitting_parm}
\end{table}
In the following, we derive the quantum buckling Hamiltonian of strained MOF-5. Assume a sufficiently high potential barrier. Then, the molecule only buckles in one potential well, e.g., the right-buckled solution. The corresponding energy is $E_g$ and can be approximated as follows. Around the classical solution $b\approx b_0$, the potential is harmonic with $V(b_0 + \delta b)\approx V_0 + 4 a b_0^2\,\delta b^2$, i.e., a quantum harmonic oscillator. The ground state energy is given by
\begin{equation}
    E_g \approx V_0 + E_0 = V_0 + \hbar\sqrt{\frac{2 a b_0^2}{m}}.
\end{equation}
The wave function $\psi_r$ is normalized as $\abs{\psi_r}^2=1$ over the right potential well. A similar construction is done for the wave function $\psi_l$ describing the buckling in the left potential well. As the barrier between both wells is finite, tunneling is allowed. We describe the effective Hamiltonian incorporating the tunneling with tunneling strength $t$ by
\begin{equation}
    \op{h} = \left(\op{a}_l^\dagger,\op{a}_r^\dagger\right)\left(
    \begin{array}{cc}
        E_g & t \\
        t & E_g
    \end{array}
    \right)\left(
    \begin{array}{c}
        \op{a}_l\\ \op{a}_r
    \end{array}\right).
    \label{bucklingham}
\end{equation}
Here, $\op{a}_l^\dagger$ and $\op{a}_r^\dagger$ ($\op{a}_l$ and $\op{a}_r$) are creation (annihilation operators) for the left-buckled and right-buckled solution, respectively. The energy levels correspond to the symmetric and antisymmetric solution of the system,
\begin{equation}
    E_\pm=E_g\pm t, \quad\op{\psi}_\pm = \frac{1}{\sqrt{2}}\left(\op{a}_l\pm\op{a}_r\right).
\end{equation}
Following Ref \cite{savel2006quantum}, we can approximate $t$ as follows
\begin{equation}
    t = \frac{2\hbar}{\pi} \sqrt{\frac{2ab_0^2}{m}}  \exp\left[
    -\frac{\pi}{2\hbar}\sqrt{\frac{m}{2ab_0^2}} (\delta V - E_0)
    \right].
\end{equation}
\begin{figure}[t!]
    \centering
    \includegraphics[width=0.45\textwidth]{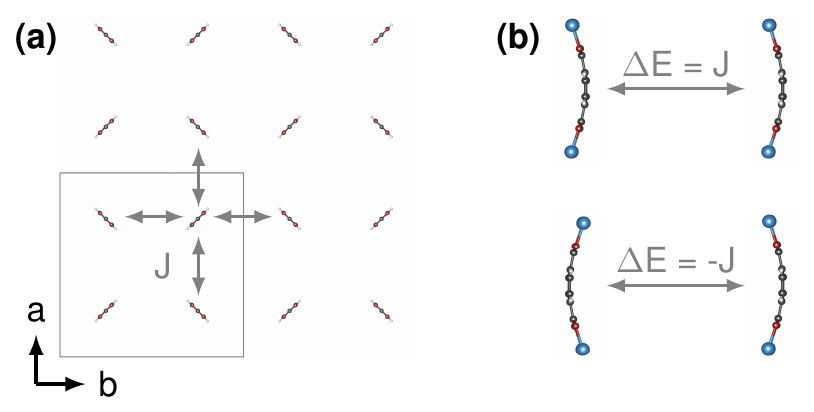}
    \caption{Exchange interaction of buckled linker molecules. (a) shows the molecular arrangement of BDC linker molecules in MOF-5 in the $\vec{a}-\vec{b}$ plane. Molecules form the pattern of a square lattice. (b) Shows the energy difference between ferro- and antiferro-buckling.}
    \label{Jcoupling}
\end{figure}
So far, we focussed on the buckling of a single molecule. However, MOF-5 is a lattice periodic framework. As a result, the buckling of adjacent molecules $i$ and $j$ is coupled, with the coupling strength $J_{ij}$. As shown in Fig. \ref{Jcoupling}, for $J_{ij}>0$ ($J_{ij}<0$) the buckling in the same (opposite) buckling state is energetically preferred. Merging this result with the single-molecule buckling Hamiltonian of \eqref{bucklingham}, we can write down the effective quantum buckling Hamiltonian for the MOF in the following way (we neglect the constant energy shift $E_0$),
\begin{equation}
    \mat{H} = -t \sum_i \mat{\sigma}^1_i - \sum_{ij} J_{ij} \mat{\sigma}^3_i \mat{\sigma}^3_j.
    \label{totalbucklingham}
\end{equation}
Equation \eqref{totalbucklingham} is the mechanical buckling version of the well-studied transverse field Ising model. The transverse field Ising model has been applied intensively to study order-disorder ferroelectrics, simple ferromagnets, simple Jahn-Teller systems and more \cite{stinchcombe1973ising,suzuki2012quantum}. In the following we summarize a few key results and their interpretation for quantum buckling.
\begin{figure*}[t!]
    \centering
    \includegraphics[width=0.9\textwidth]{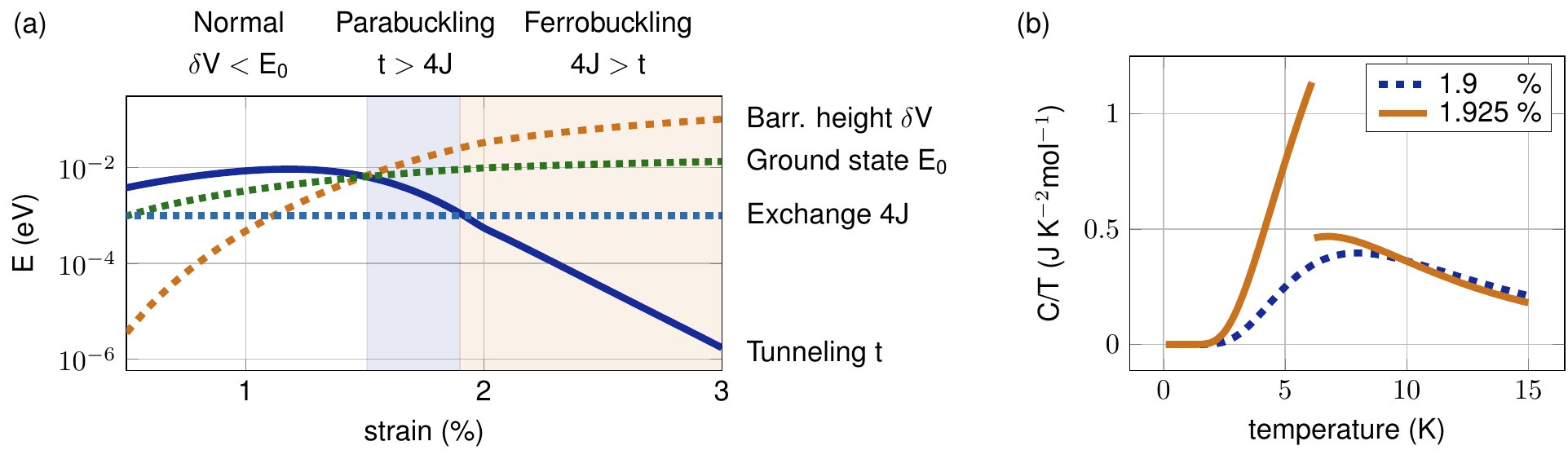}
    \caption{\change{(a) Phase diagram of strained MOF-5. If the single-well ground state energy is higher than the barrier height, no buckling occurs (normal phase). For large enough barrier height and dominating tunneling $t>4J$ molecules are in a superposition of left and right buckled states with no long-range order (parabuckling). For dominating exchange energy $4J>t$, buckled molecules form longrange order. For MOF-5, the expected transition temperature for 3\% strain is $\approx 10$~K. (b) Specific heat close to the quantum phase transition between para- and ferrobuckling. In the ferrobuckling regime, the specific heat contribution of the collective buckling state exhibits a discontinuity at the transition temperature.}}
    \label{energetics}
\end{figure*}

\change{The zero temperature phase diagram of strained MOF-5 is shown in Fig. \ref{energetics}. Depending on the strain strength, three phases are present. 
For sufficiently weak strain, the double-well potential barrier is lower than the lowest lying level of the single-well solution ($\delta V < E_0$). As a result, the quantum state of the molecule does not experience the presence of two distinct minima and the material remains in a normal state, with no buckling present. On the opposite, for sufficiently large strain and dominating exchange $\left|J_{ij}\right|\gg t$, the MOF will show ordered buckling. The simplest order possible would be the ferrobuckling state ($J_{ij}>0$), with all molecules occupying buckling states pointing in the same direction, i.e., $\left<\sigma_i^z\right> \approx 1$, $\left<\sigma_i^x \right>\approx 0$. Between the normal and the ferrobuckling phase is the parabuckling phase. Here, the energy of the single-well solutions lies below the potential barrier ($\delta V > E_0$). Also, the tunneling term for each molecule dominates the exchange between adjacent molecules ($t > 4J$), 
$\left<\sigma_i^z\right> \approx 0$, $\left<\sigma_i^x \right>\approx 1$. }

\change{In the mean-field approximation, we write \eqref{totalbucklingham} in terms of single-site contributions \cite{suzuki2012quantum},
\begin{equation}
    \op{H} \approx -\sum_{i} \vec{\op{h}}_i\cdot\vec{\op{\sigma}}_i,\qquad \vec{\op{h}}_i=t \hat{\vec{x}} + \sum_j J_{ij} \left<\op{\sigma}^3_j\right> \hat{\vec{z}}.
\end{equation}
Assuming weak variations in the buckling state $\left<\sigma_j^3\right>\approx\left<\sigma^3\right>$ and nearest-neighbor approximation, each single-site contribution $\vec{\op{h}}_i\cdot\vec{\op{\sigma}}_i$ has eigenvalues $h_\pm = \pm \sqrt{t^2+\left(4J\left<\sigma^3\right>\right)^2}=\pm\left|\vec{h}\right|$. As a result, we obtain for the buckling per site $\left<\vec{\sigma}\right> = \frac{\vec{h}}{h}\tanh\left(\beta h\right)$, with $\beta=(k_bT)^{-1}$. From the expectation value $\left<\sigma^3\right>$ we can deduce the critical temperature given by
}
\begin{equation}
    \tanh\left(\beta_c t \right) = \frac{t}{4J} \quad \xrightarrow[]{J \gg t} \quad k_B T_c \approx 4J.
    \label{TCapprox}
\end{equation}
\change{To estimate the exchange $J$ for MOF-5, we performed DFTtotal energy  calculations on a pair of molecules for the two configurations shown in Fig. \ref{Jcoupling}. The exchange $J$ is then obtained from the energy difference of the ferro- and antiferrobuckled configurations. The result was a fairly constant value of $\approx 0.25$~meV in the low strain regime. Hence, according to equation \eqref{TCapprox}, the transition temperature is $\approx 10$~K. We note that we focus on the regime of sufficiently small strain which justifies the nearest neighbor approximation. A more detailed investigation of the strength of nearest neighbor, second nearest neighbor, and four-$\sigma^3$ interactions would be necessary, but is outside the scope of this paper. In particular, the latter is expected to become significant for large strain. Such an extended four-spin transverse field Ising model is in close connection to the 8-vertex model, which has been intensively discussed \cite{baxter1971eight,baxter2016exactly}. Such a model also captures the transition between ordered and glass states. }

\change{Besides the transition temperature, mean-field approximation allows to estimate the ensemble average of the enegry, given by $\left<E\right>=-h\tanh\left(\beta h\right).$ We numerically evaluate the specific heat contribution due to collective buckling as $C = \frac{\partial \left<E\right>}{\partial T}$, taking into account the implicit temperature dependence of $\left<\sigma^3\right>$. We plot $C/T$ per mol in Fig \ref{energetics} (b) for a strain close to the quantum critical point between para- and ferrobuckling. In the parabuckling regime (dotted blue line, 1.9 \% strain) the specific heat is a smooth function in temperature. In contrast, in the ferrobuckling regime (solid orange line, 2 \% strain), a phase transition at a temparature of $\approx 6.15$~K leads to a discontinuity in the specific heat. Hence, a strain-dependent measurement of the specific heat at low temperatures could provide experimental evidence for the quantumbuckling phases. }

Collective excitations of the ordered buckling states emerge similarly to magnons in magnetically ordered systems. 
The buckleon excitations of the parabuckling phase can be estimated by evaluating the Heisenberg equation of motion to calculate $\mat{\ddot{\sigma}}^z_i$. Using the random phase approximation $\mat{\sigma}_i^x\mat{\sigma}_j^z \approx \left<\mat{\sigma}_i^x\right>\mat{\sigma}_j^z+\mat{\sigma}_i^x\left<\mat{\sigma}_j^z\right>$ \cite{suzuki2012quantum}, we obtain for the effective square lattice of MOF-5
\begin{equation}
    \mat{\ddot{\sigma}}_i^z + \omega_{\vec{q}}^2\mat{\sigma}_i^z = 0,\quad \omega_{\vec{q}} = \pm \frac{2t}{\hbar}\sqrt{1-\frac{2J}{t}\left[\cos{q_x}+\cos{q_y}\right]}.
\end{equation}
At $\vec{q}=0$, the spectrum has a gap $\Delta = 4\sqrt{t(t-4J)}$. As a result, interactions mediated by buckleons in the parabuckling phase are exponentially decaying and short ranged. At the quantum phase transition between ferro and parabuckling phase, $t\approx 4J$, the gap closes, leading to significantly enhanced interactions due to strong fluctuations. 

For example, in quantum ferroelectrics, fluctuations at the quantum critical point between the para and ferroelectric phase enhance superconductivity \cite{rowley2014ferroelectric, Edge2015}. In contrast to the ferroelectric counterparts, the quantum parabuckling-ferrobuckling phase transition can be induced straightforwardly by uniaxial pressure (the calculated bulk modulus of MOF-5 is 15.37~GPa \cite{yang2010theoretical}, whereas the bulk modulus of SrTiO$_3$ is 172.1~GPa \cite{beattie1971pressure}). \change{Modifying the formalism of Edge \textit{et al.} \cite{Edge2015} and the strong-coupling theory of McMillan \cite{mcmillan1968transition} to the present case of electron-buckleon mediated coupling, the superconducting coupling strength goes as 
\begin{align}
\lambda &= \int_0^\infty \alpha^2(\omega)F(\omega)\frac{d\omega}{\omega}\\ &\approx \alpha^2 \int \mathrm{d}\vec{q}\,\frac{1}{\frac{2t}{\hbar}\sqrt{1-\frac{2J}{t}\left[\cos{q_x}+\cos{q_y}\right]}}.
\end{align}
Here, we assumed a constant coupling constant $\alpha^2(\omega)\approx\alpha^2$ and a spectral density $F(\omega) = \int \mathrm{d}\vec{q}\,\delta(\omega-\omega_{\vec{q}})$. 
Hence, at the ferrobuckling transition where $\frac{4J}{t}\approx 1$ and $\omega_{\vec{q}=0} \rightarrow 0$,} the superconducting interaction strength is significantly enhanced. We note that MOF-5 is an insulator with a gap of $\approx 3.4$~eV \cite{yang2010theoretical}. Hence, to observe superconductivity, the material would have to be doped accordingly. \change{Currently, only very few examples of superconductivity in MOFs are known \cite{takenaka2021strongly, huang2018superconductivity}. These examples are most-likely based on strong correlations, similar to high-$T_c$ superconductors.} We believe, the emergence of order due to quantum critical buckling is a \change{different and more} general concept in MOFs. This concept goes beyond MOF-5 and superconductivity. 

In summary, we showed that applying uniaxial pressure to MOFs can lead to buckling of the organic linker molecules. At low temperatures the buckling of individual molecules is of quantum nature, where the molecule is in a superposition of left- and right-buckled states. Also, the buckling of adjacent molecules is not independent but weakly coupled. As a result, the material can undergo a phase transition into a collective para- and ferrobuckling state. MOFs are soft. Therefore, the tuning of the quantum buckling phases by uniaxial pressure can be achieved straightforwardly. The emergence of a phase transition at low temperatures should be seen in the specific heat of the material. For MOF-5 and a strain of $\approx2-5~\%$, we expect the phase transition into a ferrobuckling phase to take place at $\approx 10$~K. By slowly decreasing the uniaxial pressure to $<2~\%$ a quantum phase transition to a parabuckling phase is expected. The quantum critical point between the two phases might be a prominent experimental platform to investigate novel types of fluctuation induced order.

{\it Acknowledgment.} We are grateful to Marc Suñé Simon, Jérémy Vachier, Anthony Bonfils, Prabal Negi, Bart Olsthoorn, Vladimir Juričić, and Alexander V. Balatsky for inspiring discussions. We acknowledge support from research funding granted to A. V. Balatsky, i.e., VILLUM FONDEN via the Centre of Excellence for Dirac Materials (Grant No. 11744), the European Research Council under the European Union Seventh Framework ERS-2018-SYG
810451 HERO, the Knut and Alice Wallenberg Foundation KAW 2018.0104. 
Computational resources were provided by the Swedish National Infrastructure for Computing (SNIC) via the High Performance Computing Centre North (HPC2N) and the Uppsala Multidisciplinary Centre for Advanced Computational Science (UPPMAX).
\bibliography{references}

\begin{thebibliography}{26}%
\makeatletter
\providecommand \@ifxundefined [1]{%
 \@ifx{#1\undefined}
}%
\providecommand \@ifnum [1]{%
 \ifnum #1\expandafter \@firstoftwo
 \else \expandafter \@secondoftwo
 \fi
}%
\providecommand \@ifx [1]{%
 \ifx #1\expandafter \@firstoftwo
 \else \expandafter \@secondoftwo
 \fi
}%
\providecommand \natexlab [1]{#1}%
\providecommand \enquote  [1]{``#1''}%
\providecommand \bibnamefont  [1]{#1}%
\providecommand \bibfnamefont [1]{#1}%
\providecommand \citenamefont [1]{#1}%
\providecommand \href@noop [0]{\@secondoftwo}%
\providecommand \href [0]{\begingroup \@sanitize@url \@href}%
\providecommand \@href[1]{\@@startlink{#1}\@@href}%
\providecommand \@@href[1]{\endgroup#1\@@endlink}%
\providecommand \@sanitize@url [0]{\catcode `\\12\catcode `\$12\catcode
  `\&12\catcode `\#12\catcode `\^12\catcode `\_12\catcode `\%12\relax}%
\providecommand \@@startlink[1]{}%
\providecommand \@@endlink[0]{}%
\providecommand \url  [0]{\begingroup\@sanitize@url \@url }%
\providecommand \@url [1]{\endgroup\@href {#1}{\urlprefix }}%
\providecommand \urlprefix  [0]{URL }%
\providecommand \Eprint [0]{\href }%
\providecommand \doibase [0]{http://dx.doi.org/}%
\providecommand \selectlanguage [0]{\@gobble}%
\providecommand \bibinfo  [0]{\@secondoftwo}%
\providecommand \bibfield  [0]{\@secondoftwo}%
\providecommand \translation [1]{[#1]}%
\providecommand \BibitemOpen [0]{}%
\providecommand \bibitemStop [0]{}%
\providecommand \bibitemNoStop [0]{.\EOS\space}%
\providecommand \EOS [0]{\spacefactor3000\relax}%
\providecommand \BibitemShut  [1]{\csname bibitem#1\endcsname}%
\let\auto@bib@innerbib\@empty
\bibitem [{\citenamefont {Erbil}\ \emph {et~al.}(2020)\citenamefont {Erbil},
  \citenamefont {Hatipoglu}, \citenamefont {Yanik}, \citenamefont {Ghavami},
  \citenamefont {Ari}, \citenamefont {Yuksel},\ and\ \citenamefont
  {Hanay}}]{nanobuckling}%
  \BibitemOpen
  \bibfield  {author} {\bibinfo {author} {\bibfnamefont {S.~O.}\ \bibnamefont
  {Erbil}}, \bibinfo {author} {\bibfnamefont {U.}~\bibnamefont {Hatipoglu}},
  \bibinfo {author} {\bibfnamefont {C.}~\bibnamefont {Yanik}}, \bibinfo
  {author} {\bibfnamefont {M.}~\bibnamefont {Ghavami}}, \bibinfo {author}
  {\bibfnamefont {A.~B.}\ \bibnamefont {Ari}}, \bibinfo {author} {\bibfnamefont
  {M.}~\bibnamefont {Yuksel}}, \ and\ \bibinfo {author} {\bibfnamefont {M.~S.}\
  \bibnamefont {Hanay}},\ }\href {\doibase 10.1103/PhysRevLett.124.046101}
  {\bibfield  {journal} {\bibinfo  {journal} {Phys. Rev. Lett.}\ }\textbf
  {\bibinfo {volume} {124}},\ \bibinfo {pages} {046101} (\bibinfo {year}
  {2020})}\BibitemShut {NoStop}%
\bibitem [{\citenamefont {Savel'ev}\ \emph {et~al.}(2006)\citenamefont
  {Savel'ev}, \citenamefont {Hu},\ and\ \citenamefont
  {Nori}}]{savel2006quantum}%
  \BibitemOpen
  \bibfield  {author} {\bibinfo {author} {\bibfnamefont {S.}~\bibnamefont
  {Savel'ev}}, \bibinfo {author} {\bibfnamefont {X.}~\bibnamefont {Hu}}, \ and\
  \bibinfo {author} {\bibfnamefont {F.}~\bibnamefont {Nori}},\ }\href@noop {}
  {\bibfield  {journal} {\bibinfo  {journal} {New Journal of Physics}\ }\textbf
  {\bibinfo {volume} {8}},\ \bibinfo {pages} {105} (\bibinfo {year}
  {2006})}\BibitemShut {NoStop}%
\bibitem [{\citenamefont {Savel’ev}\ \emph {et~al.}(2007)\citenamefont
  {Savel’ev}, \citenamefont {Rakhmanov}, \citenamefont {Hu}, \citenamefont
  {Kasumov},\ and\ \citenamefont {Nori}}]{savel2007quantum}%
  \BibitemOpen
  \bibfield  {author} {\bibinfo {author} {\bibfnamefont {S.}~\bibnamefont
  {Savel’ev}}, \bibinfo {author} {\bibfnamefont {A.}~\bibnamefont
  {Rakhmanov}}, \bibinfo {author} {\bibfnamefont {X.}~\bibnamefont {Hu}},
  \bibinfo {author} {\bibfnamefont {A.}~\bibnamefont {Kasumov}}, \ and\
  \bibinfo {author} {\bibfnamefont {F.}~\bibnamefont {Nori}},\ }\href@noop {}
  {\bibfield  {journal} {\bibinfo  {journal} {Physical Review B}\ }\textbf
  {\bibinfo {volume} {75}},\ \bibinfo {pages} {165417} (\bibinfo {year}
  {2007})}\BibitemShut {NoStop}%
\bibitem [{\citenamefont {Pistolesi}\ \emph {et~al.}(2021)\citenamefont
  {Pistolesi}, \citenamefont {Cleland},\ and\ \citenamefont
  {Bachtold}}]{qubit1}%
  \BibitemOpen
  \bibfield  {author} {\bibinfo {author} {\bibfnamefont {F.}~\bibnamefont
  {Pistolesi}}, \bibinfo {author} {\bibfnamefont {A.~N.}\ \bibnamefont
  {Cleland}}, \ and\ \bibinfo {author} {\bibfnamefont {A.}~\bibnamefont
  {Bachtold}},\ }\href {\doibase 10.1103/PhysRevX.11.031027} {\bibfield
  {journal} {\bibinfo  {journal} {Phys. Rev. X}\ }\textbf {\bibinfo {volume}
  {11}},\ \bibinfo {pages} {031027} (\bibinfo {year} {2021})}\BibitemShut
  {NoStop}%
\bibitem [{\citenamefont {Li}\ \emph {et~al.}(1999)\citenamefont {Li},
  \citenamefont {Eddaoudi}, \citenamefont {O'Keeffe},\ and\ \citenamefont
  {Yaghi}}]{li1999design}%
  \BibitemOpen
  \bibfield  {author} {\bibinfo {author} {\bibfnamefont {H.}~\bibnamefont
  {Li}}, \bibinfo {author} {\bibfnamefont {M.}~\bibnamefont {Eddaoudi}},
  \bibinfo {author} {\bibfnamefont {M.}~\bibnamefont {O'Keeffe}}, \ and\
  \bibinfo {author} {\bibfnamefont {O.~M.}\ \bibnamefont {Yaghi}},\ }\href@noop
  {} {\bibfield  {journal} {\bibinfo  {journal} {nature}\ }\textbf {\bibinfo
  {volume} {402}},\ \bibinfo {pages} {276} (\bibinfo {year}
  {1999})}\BibitemShut {NoStop}%
\bibitem [{\citenamefont {Moosavi}\ \emph {et~al.}(2020)\citenamefont
  {Moosavi}, \citenamefont {Nandy}, \citenamefont {Jablonka}, \citenamefont
  {Ongari}, \citenamefont {Janet}, \citenamefont {Boyd}, \citenamefont {Lee},
  \citenamefont {Smit},\ and\ \citenamefont
  {Kulik}}]{moosavi2020understanding}%
  \BibitemOpen
  \bibfield  {author} {\bibinfo {author} {\bibfnamefont {S.~M.}\ \bibnamefont
  {Moosavi}}, \bibinfo {author} {\bibfnamefont {A.}~\bibnamefont {Nandy}},
  \bibinfo {author} {\bibfnamefont {K.~M.}\ \bibnamefont {Jablonka}}, \bibinfo
  {author} {\bibfnamefont {D.}~\bibnamefont {Ongari}}, \bibinfo {author}
  {\bibfnamefont {J.~P.}\ \bibnamefont {Janet}}, \bibinfo {author}
  {\bibfnamefont {P.~G.}\ \bibnamefont {Boyd}}, \bibinfo {author}
  {\bibfnamefont {Y.}~\bibnamefont {Lee}}, \bibinfo {author} {\bibfnamefont
  {B.}~\bibnamefont {Smit}}, \ and\ \bibinfo {author} {\bibfnamefont {H.~J.}\
  \bibnamefont {Kulik}},\ }\href@noop {} {\bibfield  {journal} {\bibinfo
  {journal} {Nature communications}\ }\textbf {\bibinfo {volume} {11}},\
  \bibinfo {pages} {1} (\bibinfo {year} {2020})}\BibitemShut {NoStop}%
\bibitem [{\citenamefont {Yuan}\ \emph {et~al.}(2018)\citenamefont {Yuan},
  \citenamefont {Feng}, \citenamefont {Wang}, \citenamefont {Pang},
  \citenamefont {Bosch}, \citenamefont {Lollar}, \citenamefont {Sun},
  \citenamefont {Qin}, \citenamefont {Yang}, \citenamefont {Zhang},
  \citenamefont {Wang}, \citenamefont {Zou}, \citenamefont {Zhang},
  \citenamefont {Zhang}, \citenamefont {Fang}, \citenamefont {Li},\ and\
  \citenamefont {Zhou}}]{mofsrev1}%
  \BibitemOpen
  \bibfield  {author} {\bibinfo {author} {\bibfnamefont {S.}~\bibnamefont
  {Yuan}}, \bibinfo {author} {\bibfnamefont {L.}~\bibnamefont {Feng}}, \bibinfo
  {author} {\bibfnamefont {K.}~\bibnamefont {Wang}}, \bibinfo {author}
  {\bibfnamefont {J.}~\bibnamefont {Pang}}, \bibinfo {author} {\bibfnamefont
  {M.}~\bibnamefont {Bosch}}, \bibinfo {author} {\bibfnamefont
  {C.}~\bibnamefont {Lollar}}, \bibinfo {author} {\bibfnamefont
  {Y.}~\bibnamefont {Sun}}, \bibinfo {author} {\bibfnamefont {J.}~\bibnamefont
  {Qin}}, \bibinfo {author} {\bibfnamefont {X.}~\bibnamefont {Yang}}, \bibinfo
  {author} {\bibfnamefont {P.}~\bibnamefont {Zhang}}, \bibinfo {author}
  {\bibfnamefont {Q.}~\bibnamefont {Wang}}, \bibinfo {author} {\bibfnamefont
  {L.}~\bibnamefont {Zou}}, \bibinfo {author} {\bibfnamefont {Y.}~\bibnamefont
  {Zhang}}, \bibinfo {author} {\bibfnamefont {L.}~\bibnamefont {Zhang}},
  \bibinfo {author} {\bibfnamefont {Y.}~\bibnamefont {Fang}}, \bibinfo {author}
  {\bibfnamefont {J.}~\bibnamefont {Li}}, \ and\ \bibinfo {author}
  {\bibfnamefont {H.-C.}\ \bibnamefont {Zhou}},\ }\href {\doibase
  https://doi.org/10.1002/adma.201704303} {\bibfield  {journal} {\bibinfo
  {journal} {Advanced Materials}\ }\textbf {\bibinfo {volume} {30}},\ \bibinfo
  {pages} {1704303} (\bibinfo {year} {2018})},\ \Eprint
  {http://arxiv.org/abs/https://onlinelibrary.wiley.com/doi/pdf/10.1002/adma.201704303}
  {https://onlinelibrary.wiley.com/doi/pdf/10.1002/adma.201704303} \BibitemShut
  {NoStop}%
\bibitem [{\citenamefont {Jiao}\ \emph {et~al.}(2019)\citenamefont {Jiao},
  \citenamefont {Seow}, \citenamefont {Skinner}, \citenamefont {Wang},\ and\
  \citenamefont {Jiang}}]{mofsrev2}%
  \BibitemOpen
  \bibfield  {author} {\bibinfo {author} {\bibfnamefont {L.}~\bibnamefont
  {Jiao}}, \bibinfo {author} {\bibfnamefont {J.~Y.~R.}\ \bibnamefont {Seow}},
  \bibinfo {author} {\bibfnamefont {W.~S.}\ \bibnamefont {Skinner}}, \bibinfo
  {author} {\bibfnamefont {Z.~U.}\ \bibnamefont {Wang}}, \ and\ \bibinfo
  {author} {\bibfnamefont {H.-L.}\ \bibnamefont {Jiang}},\ }\href@noop {}
  {\bibfield  {journal} {\bibinfo  {journal} {Materials Today}\ }\textbf
  {\bibinfo {volume} {27}},\ \bibinfo {pages} {43} (\bibinfo {year}
  {2019})}\BibitemShut {NoStop}%
\bibitem [{\citenamefont {Furukawa}\ \emph {et~al.}(2013)\citenamefont
  {Furukawa}, \citenamefont {Cordova}, \citenamefont {O’Keeffe},\ and\
  \citenamefont {Yaghi}}]{mofsrev3}%
  \BibitemOpen
  \bibfield  {author} {\bibinfo {author} {\bibfnamefont {H.}~\bibnamefont
  {Furukawa}}, \bibinfo {author} {\bibfnamefont {K.~E.}\ \bibnamefont
  {Cordova}}, \bibinfo {author} {\bibfnamefont {M.}~\bibnamefont {O’Keeffe}},
  \ and\ \bibinfo {author} {\bibfnamefont {O.~M.}\ \bibnamefont {Yaghi}},\
  }\href {\doibase 10.1126/science.1230444} {\bibfield  {journal} {\bibinfo
  {journal} {Science}\ }\textbf {\bibinfo {volume} {341}},\ \bibinfo {pages}
  {1230444} (\bibinfo {year} {2013})}\BibitemShut {NoStop}%
\bibitem [{\citenamefont {Schneemann}\ \emph {et~al.}(2014)\citenamefont
  {Schneemann}, \citenamefont {Bon}, \citenamefont {Schwedler}, \citenamefont
  {Senkovska}, \citenamefont {Kaskel},\ and\ \citenamefont
  {Fischer}}]{schneemann2014flexible}%
  \BibitemOpen
  \bibfield  {author} {\bibinfo {author} {\bibfnamefont {A.}~\bibnamefont
  {Schneemann}}, \bibinfo {author} {\bibfnamefont {V.}~\bibnamefont {Bon}},
  \bibinfo {author} {\bibfnamefont {I.}~\bibnamefont {Schwedler}}, \bibinfo
  {author} {\bibfnamefont {I.}~\bibnamefont {Senkovska}}, \bibinfo {author}
  {\bibfnamefont {S.}~\bibnamefont {Kaskel}}, \ and\ \bibinfo {author}
  {\bibfnamefont {R.~A.}\ \bibnamefont {Fischer}},\ }\href@noop {} {\bibfield
  {journal} {\bibinfo  {journal} {Chemical Society Reviews}\ }\textbf {\bibinfo
  {volume} {43}},\ \bibinfo {pages} {6062} (\bibinfo {year}
  {2014})}\BibitemShut {NoStop}%
\bibitem [{\citenamefont {Yang}\ \emph {et~al.}(2010)\citenamefont {Yang},
  \citenamefont {Vajeeston}, \citenamefont {Ravindran}, \citenamefont
  {Fjellvag},\ and\ \citenamefont {Tilset}}]{yang2010theoretical}%
  \BibitemOpen
  \bibfield  {author} {\bibinfo {author} {\bibfnamefont {L.-M.}\ \bibnamefont
  {Yang}}, \bibinfo {author} {\bibfnamefont {P.}~\bibnamefont {Vajeeston}},
  \bibinfo {author} {\bibfnamefont {P.}~\bibnamefont {Ravindran}}, \bibinfo
  {author} {\bibfnamefont {H.}~\bibnamefont {Fjellvag}}, \ and\ \bibinfo
  {author} {\bibfnamefont {M.}~\bibnamefont {Tilset}},\ }\href@noop {}
  {\bibfield  {journal} {\bibinfo  {journal} {Inorganic chemistry}\ }\textbf
  {\bibinfo {volume} {49}},\ \bibinfo {pages} {10283} (\bibinfo {year}
  {2010})}\BibitemShut {NoStop}%
\bibitem [{\citenamefont {Kresse}\ and\ \citenamefont
  {Furthm{\"u}ller}(1996)}]{vasp}%
  \BibitemOpen
  \bibfield  {author} {\bibinfo {author} {\bibfnamefont {G.}~\bibnamefont
  {Kresse}}\ and\ \bibinfo {author} {\bibfnamefont {J.}~\bibnamefont
  {Furthm{\"u}ller}},\ }\href@noop {} {\bibfield  {journal} {\bibinfo
  {journal} {Physical Review B}\ }\textbf {\bibinfo {volume} {54}},\ \bibinfo
  {pages} {11169} (\bibinfo {year} {1996})}\BibitemShut {NoStop}%
\bibitem [{\citenamefont {Sun}\ \emph {et~al.}(2015)\citenamefont {Sun},
  \citenamefont {Ruzsinszky},\ and\ \citenamefont {Perdew}}]{SunSCAN}%
  \BibitemOpen
  \bibfield  {author} {\bibinfo {author} {\bibfnamefont {J.}~\bibnamefont
  {Sun}}, \bibinfo {author} {\bibfnamefont {A.}~\bibnamefont {Ruzsinszky}}, \
  and\ \bibinfo {author} {\bibfnamefont {J.~P.}\ \bibnamefont {Perdew}},\
  }\href {\doibase 10.1103/PhysRevLett.115.036402} {\bibfield  {journal}
  {\bibinfo  {journal} {Phys. Rev. Lett.}\ }\textbf {\bibinfo {volume} {115}},\
  \bibinfo {pages} {036402} (\bibinfo {year} {2015})}\BibitemShut {NoStop}%
\bibitem [{\citenamefont {Sabatini}\ \emph {et~al.}(2013)\citenamefont
  {Sabatini}, \citenamefont {Gorni},\ and\ \citenamefont
  {de~Gironcoli}}]{SabatinirVV10}%
  \BibitemOpen
  \bibfield  {author} {\bibinfo {author} {\bibfnamefont {R.}~\bibnamefont
  {Sabatini}}, \bibinfo {author} {\bibfnamefont {T.}~\bibnamefont {Gorni}}, \
  and\ \bibinfo {author} {\bibfnamefont {S.}~\bibnamefont {de~Gironcoli}},\
  }\href {\doibase 10.1103/PhysRevB.87.041108} {\bibfield  {journal} {\bibinfo
  {journal} {Phys. Rev. B}\ }\textbf {\bibinfo {volume} {87}},\ \bibinfo
  {pages} {041108} (\bibinfo {year} {2013})}\BibitemShut {NoStop}%
\bibitem [{\citenamefont {Peng}\ \emph {et~al.}(2016)\citenamefont {Peng},
  \citenamefont {Yang}, \citenamefont {Perdew},\ and\ \citenamefont
  {Sun}}]{SCANVdW}%
  \BibitemOpen
  \bibfield  {author} {\bibinfo {author} {\bibfnamefont {H.}~\bibnamefont
  {Peng}}, \bibinfo {author} {\bibfnamefont {Z.-H.}\ \bibnamefont {Yang}},
  \bibinfo {author} {\bibfnamefont {J.~P.}\ \bibnamefont {Perdew}}, \ and\
  \bibinfo {author} {\bibfnamefont {J.}~\bibnamefont {Sun}},\ }\href {\doibase
  10.1103/PhysRevX.6.041005} {\bibfield  {journal} {\bibinfo  {journal} {Phys.
  Rev. X}\ }\textbf {\bibinfo {volume} {6}},\ \bibinfo {pages} {041005}
  (\bibinfo {year} {2016})}\BibitemShut {NoStop}%
\bibitem [{\citenamefont {Dudarev}\ \emph {et~al.}(1998)\citenamefont
  {Dudarev}, \citenamefont {Botton}, \citenamefont {Savrasov}, \citenamefont
  {Humphreys},\ and\ \citenamefont {Sutton}}]{Dudarev1998}%
  \BibitemOpen
  \bibfield  {author} {\bibinfo {author} {\bibfnamefont {S.~L.}\ \bibnamefont
  {Dudarev}}, \bibinfo {author} {\bibfnamefont {G.~A.}\ \bibnamefont {Botton}},
  \bibinfo {author} {\bibfnamefont {S.~Y.}\ \bibnamefont {Savrasov}}, \bibinfo
  {author} {\bibfnamefont {C.~J.}\ \bibnamefont {Humphreys}}, \ and\ \bibinfo
  {author} {\bibfnamefont {A.~P.}\ \bibnamefont {Sutton}},\ }\href {\doibase
  10.1103/PhysRevB.57.1505} {\bibfield  {journal} {\bibinfo  {journal} {Phys.
  Rev. B}\ }\textbf {\bibinfo {volume} {57}},\ \bibinfo {pages} {1505}
  (\bibinfo {year} {1998})}\BibitemShut {NoStop}%
\bibitem [{\citenamefont {Stinchcombe}(1973)}]{stinchcombe1973ising}%
  \BibitemOpen
  \bibfield  {author} {\bibinfo {author} {\bibfnamefont {R.}~\bibnamefont
  {Stinchcombe}},\ }\href@noop {} {\bibfield  {journal} {\bibinfo  {journal}
  {Journal of Physics C: Solid State Physics}\ }\textbf {\bibinfo {volume}
  {6}},\ \bibinfo {pages} {2459} (\bibinfo {year} {1973})}\BibitemShut
  {NoStop}%
\bibitem [{\citenamefont {Suzuki}\ \emph {et~al.}(2012)\citenamefont {Suzuki},
  \citenamefont {Inoue},\ and\ \citenamefont
  {Chakrabarti}}]{suzuki2012quantum}%
  \BibitemOpen
  \bibfield  {author} {\bibinfo {author} {\bibfnamefont {S.}~\bibnamefont
  {Suzuki}}, \bibinfo {author} {\bibfnamefont {J.-i.}\ \bibnamefont {Inoue}}, \
  and\ \bibinfo {author} {\bibfnamefont {B.~K.}\ \bibnamefont {Chakrabarti}},\
  }\href@noop {} {\emph {\bibinfo {title} {Quantum Ising phases and transitions
  in transverse Ising models}}},\ Vol.\ \bibinfo {volume} {862}\ (\bibinfo
  {publisher} {Springer},\ \bibinfo {year} {2012})\BibitemShut {NoStop}%
\bibitem [{\citenamefont {Baxter}(1971)}]{baxter1971eight}%
  \BibitemOpen
  \bibfield  {author} {\bibinfo {author} {\bibfnamefont {R.~J.}\ \bibnamefont
  {Baxter}},\ }\href@noop {} {\bibfield  {journal} {\bibinfo  {journal}
  {Physical Review Letters}\ }\textbf {\bibinfo {volume} {26}},\ \bibinfo
  {pages} {832} (\bibinfo {year} {1971})}\BibitemShut {NoStop}%
\bibitem [{\citenamefont {Baxter}(2016)}]{baxter2016exactly}%
  \BibitemOpen
  \bibfield  {author} {\bibinfo {author} {\bibfnamefont {R.~J.}\ \bibnamefont
  {Baxter}},\ }\href@noop {} {\emph {\bibinfo {title} {Exactly solved models in
  statistical mechanics}}}\ (\bibinfo  {publisher} {Elsevier},\ \bibinfo {year}
  {2016})\BibitemShut {NoStop}%
\bibitem [{\citenamefont {Rowley}\ \emph {et~al.}(2014)\citenamefont {Rowley},
  \citenamefont {Spalek}, \citenamefont {Smith}, \citenamefont {Dean},
  \citenamefont {Itoh}, \citenamefont {Scott}, \citenamefont {Lonzarich},\ and\
  \citenamefont {Saxena}}]{rowley2014ferroelectric}%
  \BibitemOpen
  \bibfield  {author} {\bibinfo {author} {\bibfnamefont {S.}~\bibnamefont
  {Rowley}}, \bibinfo {author} {\bibfnamefont {L.}~\bibnamefont {Spalek}},
  \bibinfo {author} {\bibfnamefont {R.}~\bibnamefont {Smith}}, \bibinfo
  {author} {\bibfnamefont {M.}~\bibnamefont {Dean}}, \bibinfo {author}
  {\bibfnamefont {M.}~\bibnamefont {Itoh}}, \bibinfo {author} {\bibfnamefont
  {J.}~\bibnamefont {Scott}}, \bibinfo {author} {\bibfnamefont
  {G.}~\bibnamefont {Lonzarich}}, \ and\ \bibinfo {author} {\bibfnamefont
  {S.}~\bibnamefont {Saxena}},\ }\href@noop {} {\bibfield  {journal} {\bibinfo
  {journal} {Nature Physics}\ }\textbf {\bibinfo {volume} {10}},\ \bibinfo
  {pages} {367} (\bibinfo {year} {2014})}\BibitemShut {NoStop}%
\bibitem [{\citenamefont {Edge}\ \emph {et~al.}(2015)\citenamefont {Edge},
  \citenamefont {Kedem}, \citenamefont {Aschauer}, \citenamefont {Spaldin},\
  and\ \citenamefont {Balatsky}}]{Edge2015}%
  \BibitemOpen
  \bibfield  {author} {\bibinfo {author} {\bibfnamefont {J.~M.}\ \bibnamefont
  {Edge}}, \bibinfo {author} {\bibfnamefont {Y.}~\bibnamefont {Kedem}},
  \bibinfo {author} {\bibfnamefont {U.}~\bibnamefont {Aschauer}}, \bibinfo
  {author} {\bibfnamefont {N.~A.}\ \bibnamefont {Spaldin}}, \ and\ \bibinfo
  {author} {\bibfnamefont {A.~V.}\ \bibnamefont {Balatsky}},\ }\href {\doibase
  10.1103/PhysRevLett.115.247002} {\bibfield  {journal} {\bibinfo  {journal}
  {Phys. Rev. Lett.}\ }\textbf {\bibinfo {volume} {115}},\ \bibinfo {pages}
  {247002} (\bibinfo {year} {2015})}\BibitemShut {NoStop}%
\bibitem [{\citenamefont {Beattie}\ and\ \citenamefont
  {Samara}(1971)}]{beattie1971pressure}%
  \BibitemOpen
  \bibfield  {author} {\bibinfo {author} {\bibfnamefont {A.}~\bibnamefont
  {Beattie}}\ and\ \bibinfo {author} {\bibfnamefont {G.}~\bibnamefont
  {Samara}},\ }\href@noop {} {\bibfield  {journal} {\bibinfo  {journal}
  {Journal of Applied Physics}\ }\textbf {\bibinfo {volume} {42}},\ \bibinfo
  {pages} {2376} (\bibinfo {year} {1971})}\BibitemShut {NoStop}%
\bibitem [{\citenamefont {McMillan}(1968)}]{mcmillan1968transition}%
  \BibitemOpen
  \bibfield  {author} {\bibinfo {author} {\bibfnamefont {W.}~\bibnamefont
  {McMillan}},\ }\href@noop {} {\bibfield  {journal} {\bibinfo  {journal}
  {Physical Review}\ }\textbf {\bibinfo {volume} {167}},\ \bibinfo {pages}
  {331} (\bibinfo {year} {1968})}\BibitemShut {NoStop}%
\bibitem [{\citenamefont {Takenaka}\ \emph {et~al.}(2021)\citenamefont
  {Takenaka}, \citenamefont {Ishihara}, \citenamefont {Roppongi}, \citenamefont
  {Miao}, \citenamefont {Mizukami}, \citenamefont {Makita}, \citenamefont
  {Tsurumi}, \citenamefont {Watanabe}, \citenamefont {Takeya}, \citenamefont
  {Yamashita} \emph {et~al.}}]{takenaka2021strongly}%
  \BibitemOpen
  \bibfield  {author} {\bibinfo {author} {\bibfnamefont {T.}~\bibnamefont
  {Takenaka}}, \bibinfo {author} {\bibfnamefont {K.}~\bibnamefont {Ishihara}},
  \bibinfo {author} {\bibfnamefont {M.}~\bibnamefont {Roppongi}}, \bibinfo
  {author} {\bibfnamefont {Y.}~\bibnamefont {Miao}}, \bibinfo {author}
  {\bibfnamefont {Y.}~\bibnamefont {Mizukami}}, \bibinfo {author}
  {\bibfnamefont {T.}~\bibnamefont {Makita}}, \bibinfo {author} {\bibfnamefont
  {J.}~\bibnamefont {Tsurumi}}, \bibinfo {author} {\bibfnamefont
  {S.}~\bibnamefont {Watanabe}}, \bibinfo {author} {\bibfnamefont
  {J.}~\bibnamefont {Takeya}}, \bibinfo {author} {\bibfnamefont
  {M.}~\bibnamefont {Yamashita}},  \emph {et~al.},\ }\href@noop {} {\bibfield
  {journal} {\bibinfo  {journal} {Science Advances}\ }\textbf {\bibinfo
  {volume} {7}},\ \bibinfo {pages} {eabf3996} (\bibinfo {year}
  {2021})}\BibitemShut {NoStop}%
\bibitem [{\citenamefont {Huang}\ \emph {et~al.}(2018)\citenamefont {Huang},
  \citenamefont {Zhang}, \citenamefont {Liu}, \citenamefont {Yu}, \citenamefont
  {Chen}, \citenamefont {Xu},\ and\ \citenamefont
  {Zhu}}]{huang2018superconductivity}%
  \BibitemOpen
  \bibfield  {author} {\bibinfo {author} {\bibfnamefont {X.}~\bibnamefont
  {Huang}}, \bibinfo {author} {\bibfnamefont {S.}~\bibnamefont {Zhang}},
  \bibinfo {author} {\bibfnamefont {L.}~\bibnamefont {Liu}}, \bibinfo {author}
  {\bibfnamefont {L.}~\bibnamefont {Yu}}, \bibinfo {author} {\bibfnamefont
  {G.}~\bibnamefont {Chen}}, \bibinfo {author} {\bibfnamefont {W.}~\bibnamefont
  {Xu}}, \ and\ \bibinfo {author} {\bibfnamefont {D.}~\bibnamefont {Zhu}},\
  }\href@noop {} {\bibfield  {journal} {\bibinfo  {journal} {Angewandte Chemie
  International Edition}\ }\textbf {\bibinfo {volume} {57}},\ \bibinfo {pages}
  {146} (\bibinfo {year} {2018})}\BibitemShut {NoStop}%
\end{thebibliography}%

\end{document}